\documentclass[aip,apl,reprint]{revtex4-1}

\usepackage{graphicx}
\usepackage{epstopdf}
\usepackage{dcolumn}
\usepackage{bm}
\usepackage{amsmath}
\usepackage{amssymb}
\usepackage{wasysym}
\usepackage{color}
\usepackage{amssymb}
\usepackage{dsfont}

\begin{document}

\title{Nonlinear thermal transport and negative differential thermal conductance in graphene nanoribbons} 

\author{Jiuning Hu}
\email[]{hu49@purdue.edu}
\affiliation{School of Electrical and Computer Engineering, Purdue University, West Lafayette, Indiana 47907,USA}
\affiliation{Birck Nanotechnology Center, Purdue University, West Lafayette, Indiana 47907,USA}

\author{Yan Wang}
\affiliation{School of Mechanical Engineering, Purdue University, West Lafayette, Indiana 47907,USA}

\author{Ajit Vallabhaneni}
\affiliation{School of Mechanical Engineering, Purdue University, West Lafayette, Indiana 47907,USA}

\author{Xiulin Ruan}
\affiliation{School of Mechanical Engineering, Purdue University, West Lafayette, Indiana 47907,USA}
\affiliation{Birck Nanotechnology Center, Purdue University, West Lafayette, Indiana 47907,USA}

\author{Yong P. Chen}
\email[]{yongchen@purdue.edu}
\affiliation{Department of Physics, Purdue University, West Lafayette, Indiana 47907,USA}
\affiliation{Birck Nanotechnology Center, Purdue University, West Lafayette, Indiana 47907,USA}
\affiliation{School of Electrical and Computer Engineering, Purdue University, West Lafayette, Indiana 47907,USA}


\begin{abstract}
We employ classical molecular dynamics to study the nonlinear thermal transport in graphene nanoribbons (GNRs). For GNRs under large temperature biases beyond linear response regime, we have observed the onset of negative differential thermal conductance (NDTC). NDTC is tunable by varying the manner of applying the temperature biases. NDTC is reduced and eventually disappears when the length of the GNR increases. We have also observed NDTC in triangular GNRs, where NDTC exists only when the heat current is from the narrower to the wider end. These effects may be useful in nanoscale thermal managements and thermal signal processing utilizing GNRs.
\end{abstract}

\maketitle

Graphene,\cite{Geim07,Castro08} an atomic monolayer of graphite, has emerged as one of the most interesting materials in condensed matter physics and nanotechnology. Besides its unusual electronic properties,\cite{Castro08} graphene also has unique thermal properties, e.g., high thermal conductivities ($\sim$600-5000 W/m-K).\cite{Alexander08, Cai10, Faugeras10, Luis10, Seol10} Graphene nanoribbons (GNRs) are promising in many applications, such as their electronic band-gap tunability\cite{Han07} and edge chirality dependent thermal transport.\cite{Hu09} So far, little attention has been paid to nonlinear thermal transport in GNRs, though these nonlinear effects have been explored in ideal atomic chains,\cite{Li06,Zhong09,He09,Pereira10,He10} molecular junctions\cite{Segal06} and quantum dots.\cite{Kuo10} Here, we demonstrate negative differential thermal conductance (NDTC) in GNRs. Analogous to the electronic counterpart,\cite{Esaki58} NDTC is a useful ingredient in developing GNR-based thermal management and signal manipulation devices, such as the thermal amplifiers\cite{Li06} and thermal logic gates.\cite{Wang07}

\begin{figure}
  \includegraphics[width=0.48\textwidth]{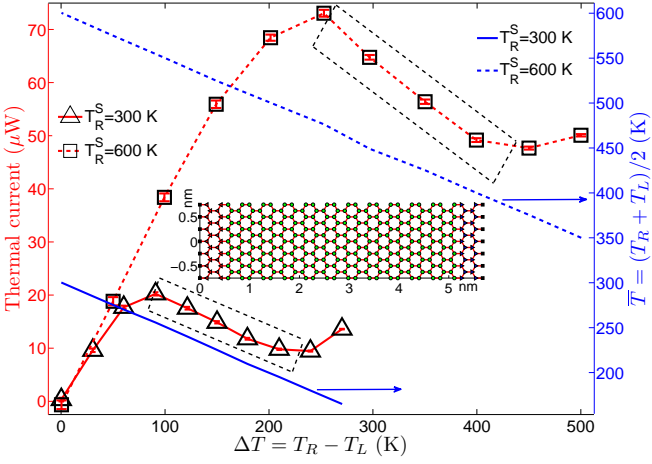}\\
  \caption{Thermal current (left vertical axis) and average temperature (right vertical axis) vs. temperature difference 
  $\Delta T$. The dashed boxes highlight NDTC. The inset shows the structure of the GNR ($\sim\text{1.5 nm}\times\text{6 nm}$). $\blacksquare$ denotes fixed boundary atoms. 
$\color{red}\blacktriangleleft$ ($\color{blue}\blacktriangleright$) denotes atoms in the left (right) thermostat. $\color{green}\CIRCLE$ denotes the remain atoms in the bulk.}
\label{symndtcstr}
\end{figure}

We study the thermal transport in GNRs using classical molecular dynamics (MD) simulations. The many-body empirical Brenner potential\cite{Brenner90} is employed to describe the carbon-carbon interactions. This method have been applied in many graphene-based systems.\cite{Hu09,Wang09,Ong10,Hu10} The structures of GNRs in this study are shown in the inset (rectangular GNR) of Fig.~\ref{symndtcstr} and the inset (triangular GNR) of Fig.~\ref{asy}. The atoms denoted by squares are fixed in position, while those denoted by left- and right-pointing triangles are placed in two Nos\'e-Hoover\cite{Nose84,Hoover85} thermostats at different temperatures $T_L$ and $T_R$, respectively. The equations of motion for atoms without position being fixed are:
\begin{equation}
\frac{d}{dt}\mathbf p_i=\mathbf F_i-\gamma_i\mathbf p_i
\end{equation}
where $\mathbf p_i$ is the momentum of the $i$-th atom, $\mathbf F_i$ is the total force acting on the $i$-th atom, and $\gamma_i$ is the Nos\'e-Hoover dynamic parameter. For the atoms denoted by circles, $\gamma_i\equiv 0$, and it recovers the NVE (constant number of atoms, volume, and energy) ensemble. For the atoms in the left and right thermostats, $\gamma_i$ obeys the equation 
\begin{equation}
\begin{aligned}
\frac{d}{dt}\gamma_i=\frac{\left[\frac{2}{3N_{L(R)}k_B}\sum_{i\in L(R)}\frac{\mathbf p_i^2}{2m}\right]-T_{L(R)}}{\tau^2T_{L(R)}},
\end{aligned}
\end{equation}
where $\tau$ is the thermostat relaxation time, 
$N_{L(R)}$ is the number of atoms in the thermostat, $k_B$  is the Boltzmann constant and $m$ is the mass of the carbon atom. More details on our numerical calculation method can be found elsewhere.\cite{Hu09,Hu09b}

First, we study the thermal transport in a rectangular GNR with armchair top and bottom edges shown in the inset of Fig.~\ref{symndtcstr} (we have obtained qualitatively similar conclusions for GNRs with zigzag edges). Since the GNR is symmetrical, we only consider $T_L\le T_R$ and define the temperature difference $\Delta T\equiv T_R-T_L$. The temperature $T_R$ is kept as a constant. As we can see from both curves in Fig.~\ref{symndtcstr}, for small temperature difference (e.g., $\Delta T<60$ K for $T_R=300$ K and $\Delta T<150$ K for $T_R=600$ K), the thermal current increases approximately linearly as $\Delta T$ increases, as expected from Fourier's law. Interestingly, for some range of higher $\Delta T$, the thermal current decreases as $\Delta T$ increases (the dashed boxes in Fig.~\ref{symndtcstr}), indicating the onset of NDTC. It is a reasonable approximation to consider thermal current as proportional to the product of thermal conductivity $\kappa$ of the GNR and $\Delta T$. Our previous study\cite{Hu09} has shown that $\kappa$ increases with the average temperature $\overline{T}\equiv(T_L+T_R)/2=T_R-\Delta T/2$. We have plotted $\overline{T}$ (labeled at the right vertical axis and indicated by right-pointing arrows for Fig.~1-3 and in the subplot of Fig.~\ref{trans}(b)) as a function of $\Delta T$ in all figures. Since $\overline{T}$ decreases with $\Delta T$, $\kappa$ decreases with increasing $\Delta T$. The resulting trend of the thermal current as a function of $\Delta T$ is thus a competition between decreasing $\kappa$ and increasing $\Delta T$. In the $\Delta T$ range displaying NDTC, the decrease of $\kappa$ with $\Delta T$ dominates. 
We have found that there is no NDTC (shown in Fig.~\ref{trans}) if $T_L$ is larger than the constant $T_R$, i.e., if $\overline{T}$ increases with $\Delta T$ (thus without the above competition). Note that for large $\Delta T$ beyond linear response, strictly speaking thermal conductivity is not well defined. Thus, in the above explanation, $\kappa$ is considered to be an effective, average thermal conductivity. Similar arguments have been applied in analysing thermal transport in 1D atomic chains.\cite{He10} 

\begin{figure}
  \includegraphics[width=0.48\textwidth]{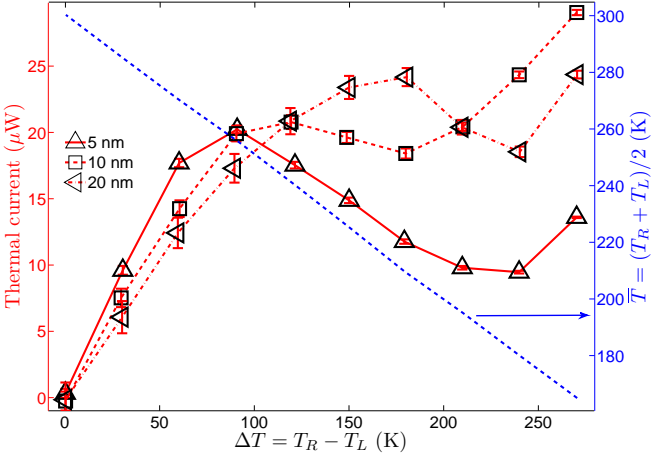}\\
  \caption{Thermal current (left vertical axis) and average temperature (right vertical axis) vs. temperature difference $\Delta T$ in GNRs with the similar structure as the GNR in the inset of Fig.~\ref{symndtcstr}, except for different lengths. In all these plots, $T_R=300$ K and $T_L$ is varied from 300 K to 30 K.}\label{symsize}
\end{figure}

Second, we study the length dependence of NDTC in GNRs. For all three GNRs of different lengths in Fig.~\ref{symsize}, $T_R=300$ K while $T_L$ is varied from $T_R$ to $30$ K. As the GNR length is increased, the $\Delta T$ value for the onset of NDTC increases and the $\Delta T$ range where NDTC exists shrinks. We thus suggest that NDTC will eventually disappear if the length of GNR exceeds some critical value. We have verified this using LAMMPS package\cite{Plimpton95} and velocity scaling\cite{Huang06} MD, and found no NDTC in a 50 nm long GNR with similar structure as that studied in Fig.~\ref{symndtcstr}.

\begin{figure}
  \includegraphics[width=0.48\textwidth]{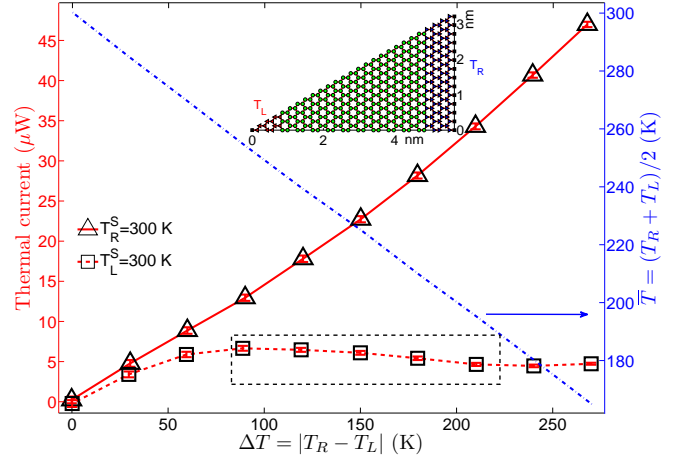}\\
  \caption{Thermal current (left vertical axis) and average temperature (right vertical axis) vs. temperature difference $\Delta T$ in triangular GNR shown in the inset. The labels for the GNR structure have the same meaning as that in the inset in Fig.~\ref{symndtcstr}. The dashed box highlights NDTC.}\label{asy}
\end{figure}

Besides these nonlinear effects in symmetrical GNRs, we also explore the possibility of NDTC in an asymmetrical triangular GNR, shown in the inset of Fig.~\ref{asy}. Our previous study has pointed out that thermal rectification exists in this asymmetrical GNR.\cite{Hu09}
As we see from Fig.~\ref{asy}, here the nonlinear thermal transport is also direction-dependent. NDTC appears when the temperature of the narrower end is held at $T_L=300$ K and the temperature $T_R$ of the wider end is varied from 300 K to 30 K (solid line in Fig.~\ref{asy}). However, there is no NDTC when the values of $T_L$ and $T_R$ are interchanged (dashed line in Fig.~\ref{asy}). This provides another possibility to control the nonlinear thermal transport and NDTC in GNRs by engineering the shape of GNRs.

\begin{figure}
  \includegraphics[width=0.48\textwidth]{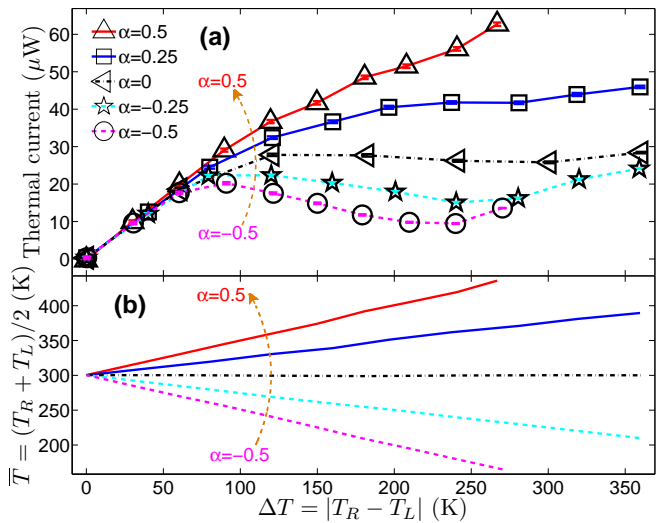}\\
  \caption{Thermal current (a) and average temperature (b) vs. temperature difference $\Delta T$ for different values of $\alpha$ for the GNR shown in the inset of Fig.~\ref{symndtcstr}. Note that $\alpha=0.5$ (-0.5) corresponds to $T_{L(R)}$ fixed at 300 K while $T_{R(L)}$ is varied.
}\label{trans}
\end{figure}

In general, the way to tune the thermal current in the two-terminal thermal devices is very different from that in any two-terminal electronic devices. In the latter case, only the voltage difference matters. However, in thermal devices, the average temperature $\overline{T}$ is as important as the temperature difference $\Delta T$ in controlling the thermal current. For example, consider $\overline{T}=\alpha\Delta T+T_0$ with constants $\alpha$ and $T_0$, and we have $T_L=(\alpha-\frac{1}{2})\Delta T+T_0$ and $T_R=(\alpha+\frac{1}{2})\Delta T+T_0$. 
The thermal currents and average temperature $\overline{T}$ as a function of $\Delta T$ are plotted in Fig.~\ref{trans} for the rectangular GNR shown in the inset in Fig.~\ref{symndtcstr}, where $T_0=300$ K and $\alpha$ is tuned from -0.5 to 0.5 (indicated by the dashed curved arrow in Fig.~\ref{trans}). The solid curve in Fig.~\ref{symndtcstr} corresponds to $\alpha=-0.5$. For small temperature difference in the linear response regime, the slope of thermal current vs. $\Delta T$ is independent of $\alpha$. In the nonlinear response regime (large $\Delta T$), the system transitions from a regime with NDTC to a regime without NDTC when $\alpha$ is tuned from negative to positive values. We can see a strong correlation between the the trend of the thermal current and that of the average temperature for different values of $\alpha$ in the range of $\Delta T$ from 100 K to 250 K where NDTC occurs for negative $\alpha$. For negative $\alpha$, since $\overline{T}$ decreases with $\Delta T$, the effective $\kappa$ decreases with $\Delta T$, and the occurrence of NDTC can be similarly explained as that for Fig.~\ref{symndtcstr}.

There are two independent parameters to control the thermal transport in two-terminal devices, either $(T_L,T_R)$ or $(\Delta T,\overline{T})$. Two-terminal thermal devices are actually analogous to three-terminal electronic devices. In the language of electronic transport of field effect transistors (FETs), $\Delta T$ plays the role of the drain-source voltage difference in FETs, while $\alpha$ plays the role of the gate voltage. Fig.~\ref{trans} shows the ability to realize the FET-like behaviour in GNRs.

In summary, we have studied the nonlinear thermal transport in rectangular and triangular GNRs under large temperature biases. We find that in short ($\sim$ 6 nm) rectangular GNRs the NDTC exists in a certain range of applied temperature difference. As the length of the rectangular GNR increases, NDTC gradually weakens. In triangular GNRs, NDTC only exists in the thermal current direction from the narrower to the wider end. The ability to tune and control NDTC by temperature parameters and GNR shapes provides potential ways to manage heat and manipulate thermal signals at the nanoscale.

This work is partially supported by the Semiconductor Research
Corporation (SRC) - Nanoelectronics Research Initiative (NRI) via
Midwest Institute for Nanoelectronics Discovery (MIND) and the Cooling Technologies Research Center (CTRC).


\providecommand{\noopsort}[1]{}\providecommand{\singleletter}[1]{#1}%

\end{document}